\documentclass[runningheads]{llncs}
\usepackage[T1]{fontenc}

\usepackage[sort]{cite}
\usepackage{graphicx}
\usepackage{array}
\usepackage{tabularx}
\usepackage{subfig}
\usepackage[normalem]{ulem}

\begin{document}
\title{LLMs to Support K–12 Teachers in Culturally Relevant Pedagogy: An AI Literacy Example}

\titlerunning{Exploring LLM-Supported K-12 Culturally Relevant Pedagogy}

%

\author{
  Jiayi Wang\inst{1}\orcidID{0009-0003-1665-9360} \and
  Ruiwei Xiao\inst{2}\orcidID{0000-0002-6461-7611} \and
  Xinying Hou\inst{3}\orcidID{0000-0002-1182-5839} \and
  Hanqi Li\inst{4}\orcidID{0009-0007-0659-1998} \and
  Ying Jui Tseng\inst{2}\orcidID{0009-0006-1801-6061} \and
  John Stamper\inst{2}\orcidID{0000-0002-2291-1468} \and
  Ken Koedinger\inst{2}\orcidID{0000-0002-5850-4768}
}

\authorrunning{J. Wang et al.}

\institute{
  Northwestern University, Evanston, IL 60208, USA\\
  \email{jiayiwang2025@u.northwestern.edu}
  \and
  Carnegie Mellon University, Pittsburgh, PA 15213, USA\\
  \email{\{ruiweix,\,yingjuit,\,jstamper,\,koedinger\}@cmu.edu}
  \and
  University of Michigan, Ann Arbor, MI 48109, USA\\
  \email{xyhou@umich.edu}
  \and
  New York University, New York, NY 10003, USA\\
  \email{hl4893@nyu.edu}
}

\maketitle              
\vspace{-6mm}
\begin{abstract}
Culturally Relevant Pedagogy (CRP) is vital in K-12 education, yet teachers struggle to implement CRP into practice due to time, training, and resource gaps. This study explores how Large Language Models (LLMs) can address these barriers by introducing CulturAIEd, an LLM tool that assists teachers in adapting AI literacy curricula to students’ cultural contexts. Through an exploratory pilot with four K-12 teachers, we examined CulturAIEd’s impact on CRP integration. Results showed CulturAIEd enhanced teachers’ confidence in identifying opportunities for cultural responsiveness in learning activities and making culturally responsive modifications to existing activities. They valued CulturAIEd's streamlined integration of student demographic information, immediate actionable feedback, which could result in high implementation efficiency. This exploration of teacher-AI collaboration highlights how LLM can help teachers include CRP components into their instructional practices efficiently, especially in global priorities for future-ready education, such as AI literacy.
\vspace{-3mm}
\end{abstract}
\vspace{-10mm}
\section{Introduction}
\vspace{-3mm}
Culturally Relevant Pedagogy (CRP), rooted in Ladson-Billings’ foundational framework, emphasizes centering students’ cultural identities as assets in education \cite{ladson-billingsTheoryCulturallyRelevant1995}. CRP has been successfully implemented in traditional subjects such as mathematics, history, and language arts \cite{aronsonTheoryPracticeCulturally2016a,hammondCulturallyResponsiveTeaching2015}. Despite benefits like improved engagement, achievement, and self-esteem \cite{byrdDoesCulturallyRelevant2016,deeCausalEffectsCultural2017}, teachers struggle with CRP due to limited time, training, and discomfort with cultural topics \cite{samuels2018exploring}. Concurrently, the rise of artificial intelligence (AI) in society has spurred calls for K-12 AI literacy initiatives \cite{tseng2024activeai}. However, there is a notable lack of culturally responsive AI literacy learning resources that address the needs of a diverse population of learners. This gap exists because AI literacy is an emerging subject with limited educational resources tailored to different cultural and demographic contexts \cite{drugaLandscapeTeachingResources2022,longWhatAILiteracy2020a,li2025unseen}. This dual challenge, integrating CRP into AI education, presents a critical opportunity for innovation. Large Language Models (LLMs) have shown promise in streamlining lesson planning and content adaptation \cite{royChatGPTLessonPreparation}, but their potential to support CRP remains underexplored. 

To address this gap, we present CulturAIEd, an LLM-powered system designed to help teachers contextualize AI literacy activities through CRP principles. CulturAIEd combines demographic characteristics of the students, rubric-based feedback, and just-in-time coaching to simplify the adaptation process while encouraging meaningful cultural integration. This paper investigates two research questions:
\textbf{RQ1:} How do LLM-supported tools like CulturAIEd influence teachers’ ability to design culturally responsive AI literacy activities?
\textbf{RQ2:} What are the perceived strengths and opportunities of such tools in addressing CRP implementation barriers? Through a pilot with four K–12 teachers, conducted in preparation for future mixed-methods study, we explored how CulturAIEd influences their confidence, efficiency, and pedagogical strategies. Our findings highlight the tool’s role in reducing planning burdens and fostering deeper engagement with CRP. To our knowledge, this work represents the first effort to empower K-12 educators teaching traditional subjects to design culturally responsive AI literacy activities with low effort using LLMs. By situating AI as a partner in CRP, this work advances discussions on equitable technology integration and teacher empowerment in diverse classrooms.
\vspace{-4mm}
\section{Related Work}
\vspace{-3mm}

Culturally Relevant Pedagogy (CRP) emphasizes leveraging students' cultural backgrounds as key assets in education. Ladson-Billings' foundational work defines CRP through promoting academic success, developing cultural competence, and fostering sociopolitical consciousness \cite{ladson-billingsTheoryCulturallyRelevant1995}. Empirical evidence consistently highlights CRP's positive impact on academic achievement, student engagement, attendance, motivation, social-emotional development and ethnic-racial identity \cite{johnsonCulturallyRelevantPedagogy2020,deeCausalEffectsCultural2017,byrdDoesCulturallyRelevant2016}. Nonetheless, teachers often encounter significant challenges implementing CRP, including constraints like limited time, curricular rigidity, lack of culturally relevant resources, and discomfort addressing sensitive cultural topics \cite{samuels2018exploring,mette2016teachers}. Building upon CRP, Geneva Gay introduced the closely aligned concept of Culturally Responsive Teaching (CRT), emphasizing more explicit strategies for operationalizing cultural responsiveness within daily teaching practices \cite{gayCulturallyResponsiveTeaching2000}. CRT particularly informs practical methodologies related to modifying course materials, which is perceived as challenging by many teachers. Teachers frequently require administrative support, professional training, and accessible tools to effectively integrate CRP/CRT methods. 

As AI becomes integral to societal infrastructure, there is increasing recognition of the necessity for K-12 AI literacy education \cite{casal-oteroAILiteracyK122023a}. However, integrating AI into curricula remains exploratory and lacks systematic frameworks \cite{casal-oteroAILiteracyK122023a}. Current AI literacy programs largely employ a generalized approach, minimally tailored to specific cultural contexts, despite evidence supporting the effectiveness of contextualized content \cite{eguchiContextualizingAIEducation2021}. Scholars advocate for modular, culturally adaptable AI curricula to better meet the diverse backgrounds and needs of students \cite{casal-oteroAILiteracyK122023a}. 

LLMs have emerged as promising tools to alleviate challenges in implementing CRP/CRT practices. LLMs support teachers by significantly reducing time spent on lesson planning and content adaptation \cite{10.1007/978-3-031-64315-6_49}. Early studies indicate that teachers can utilize LLMs to create lesson materials and adapt existing curricula, addressing key barriers like limited resources and heavy workloads \cite{royChatGPTLessonPreparation,baytak2024content}. However, most AI-driven teacher tools to date focus on general lesson planning or content generation rather than explicitly supporting cultural relevance. Some tools address content management, sustainability goals, or improving lesson plans, reflecting a growing interest in AI for education \cite{moorhousePreserviceLanguageTeachers2025}, but without directly incorporating CRP frameworks. CulturAIEd is novel in that it integrates a CRT checklist \cite{CulturallyResponsiveTeaching2019} and demographic customization into the LLM’s generative process, providing feedback layered with cultural pedagogical guidance. We describe CulturAIEd’s design and a pilot study evaluating its use below.
\vspace{-4mm}
\section{Methods}
\vspace{-4mm}

\begin{figure}[hbpt]
    \vspace{-4mm}
      \centering
      \includegraphics[width=0.8\linewidth]{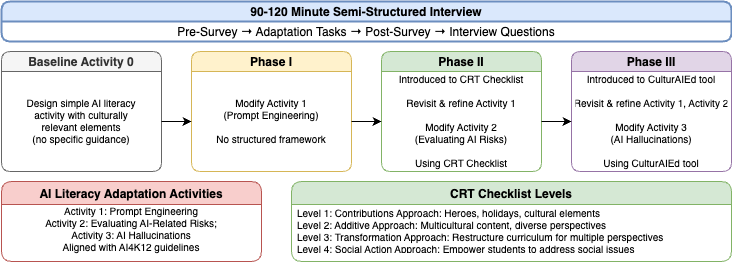}
      \vspace*{-2mm}
      \caption{Study Design}
    \vspace{-6mm}
\end{figure}

\vspace{-6mm}
\subsection{CulturAIEd Tool Design}
\vspace{-1mm}
CulturAIEd integrates principles from CRP/CRT to facilitate inclusive and effective lesson planning. Central to CulturAIEd's design is an LLM feature that provides culturally relevant content, such as examples, analogies, or contextual details, tailored specifically to the demographics of a teacher’s classroom. This function supports educators in overcoming common barriers to CRT implementation by providing concrete, relatable connections that authentically reflect students' cultural backgrounds and lived experiences, thus fostering greater engagement and deeper academic understanding. Additionally, CulturAIEd provides two tiers of lesson modifications, basic and advanced, based on the CRT checklist \cite{CulturallyResponsiveTeaching2019}, guiding teachers in culturally responsive adaptation. On the one hand, it provides teachers with flexible options suited to various instructional contexts or constraints, on the other hand, by explicitly showcasing differences between tokenistic and meaningful cultural integration, it encourages educators to move beyond surface-level adjustments towards genuine and substantive curricular changes. Complementing this, CulturAIEd employs rubric-based scoring and AI-driven feedback to evaluate and enhance lesson quality systematically. Teachers receive formative feedback highlighting areas for improvement, thereby promoting continuous professional growth through explicit and actionable recommendations. Finally, an interactive LLM chatbot provides just-in-time pedagogical support, simulating instructional coaching to address teachers' immediate needs. By embedding such support within teachers' routine planning processes, CulturAIEd advances educators’ competencies in delivering culturally responsive instruction, ultimately enriching student learning experiences. Before the interviews, we built a demo with streamlit \cite{StreamlitFasterWay2021} and gpt-4o-mini-2024-07-18 model. 

\begin{figure}[hbpt]%
    \vspace{-2mm}
    \centering
    \subfloat[\centering Design]{{\includegraphics[width=0.4\linewidth]{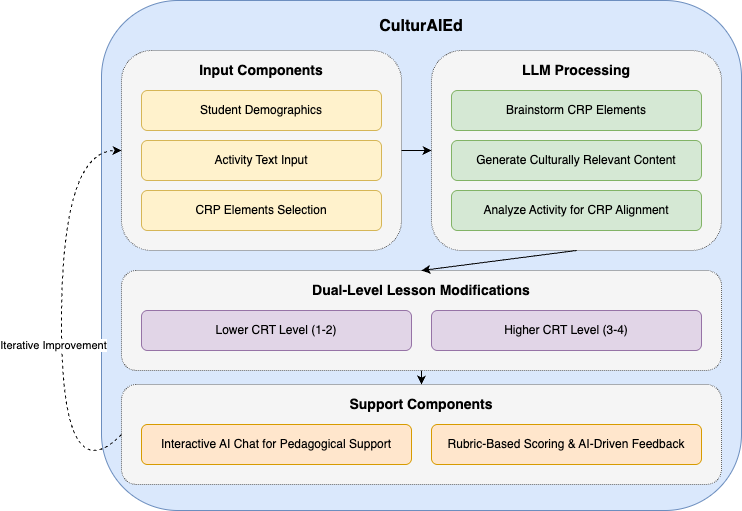} }}%
    \qquad
    \subfloat[\centering Demo]{{\includegraphics[width=0.45\linewidth]{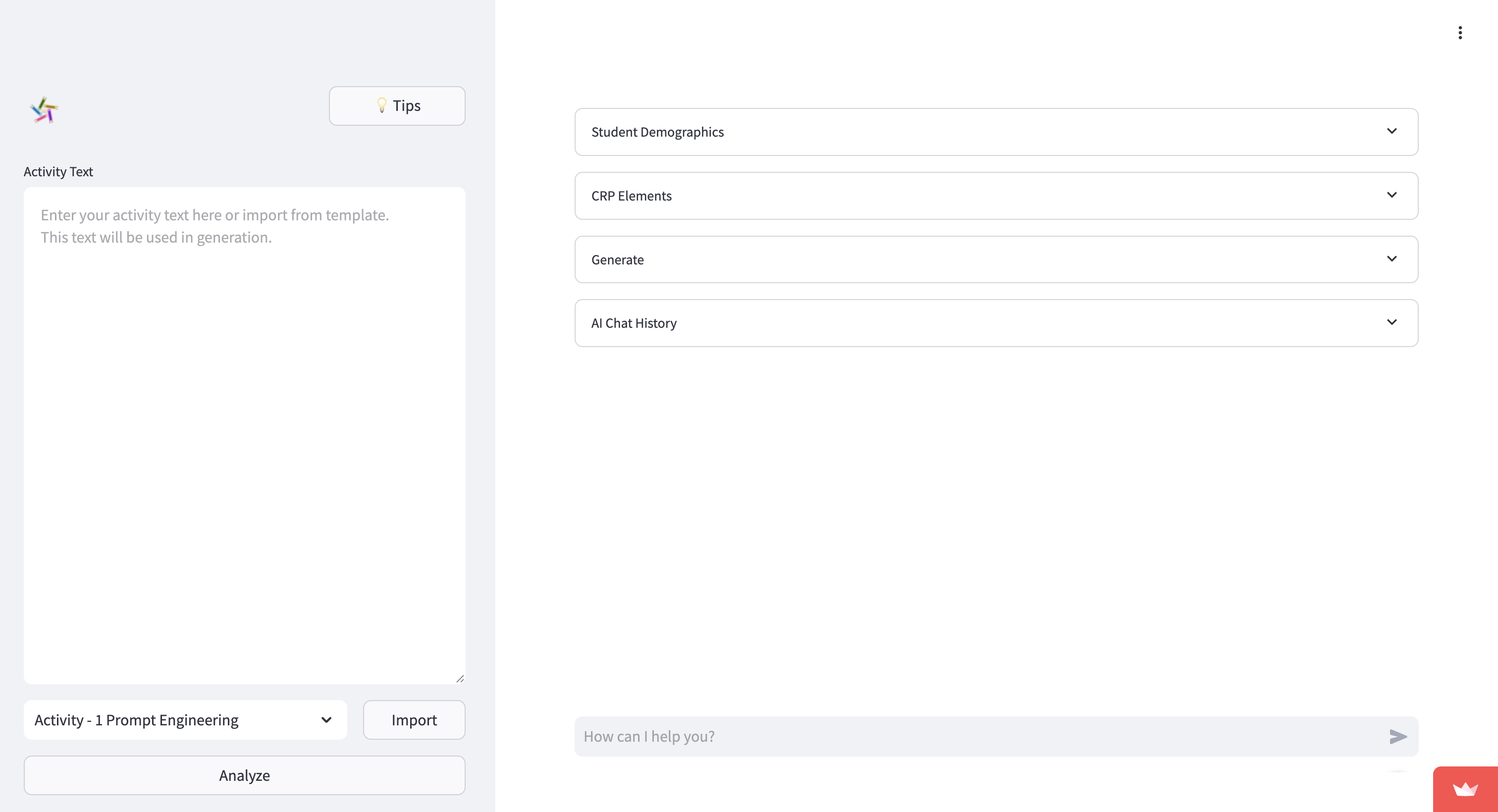} }}%
    \caption{CulturAIEd Design}%
    \vspace{-8mm}
    \label{fig:CulturAIEd_Design}%
\end{figure}

\vspace{-4mm}

\subsection{Study Procedure}

In this IRB (Institutional Review Board)-approved study, participants completed 90-120 minute semi-structured 1-on-1 Zoom sessions involving a pre-survey, adaptation tasks, post-survey, and interviews. The tasks focused on adapting AI literacy activities, aligned with AI4K12 guidelines \cite{AI4K12}, to include CRP elements. Participants started with completing \textbf{Activity 0} (designing a basic AI literacy lesson without guidance) as a baseline to observe CRP integration strategies. Participants then engaged in a structured adaptation process across three phases. In Phase I, they independently adapted a prompt engineering activity (\textbf{Activity 1}) without any guidance. In Phase II, participants employed a structured CRT Checklist \cite{CulturallyResponsiveTeaching2019} to adapt the previously addressed prompt engineering activity along with an additional task on evaluating AI-related risks (\textbf{Activity 2}), progressing through defined CRT levels: Contributions, Additive, Transformation, and Social Action. In Phase III, participants utilized CulturAIEd to refine their prior adaptations and to modify a new activity focused on AI hallucinations (\textbf{Activity 3}), exploring deeper and more meaningful integration of CRP strategies. Participants had access to generative AI tools (e.g., ChatGPT) and internet resources throughout.

\vspace{-4mm}

\subsection{Participants \& Data Analysis}

We recruited three secondary teachers from North America (P1, P2, P4) via Prolific \cite{ProlificEasilyCollect} and one from East Asia (P3), using quota sampling to ensure racial and ethnic diversity. Participants included P1 (White female, science teacher, grades 9-12, virtual public school), P2 (Black male, music education teacher, grades 9-12, public school), P3 (Asian female, information technology teacher, grade 9, public school), and P4 (White female, science teacher, grades 9-12, public school). All participants had at least 4 years of teaching experience and varied AI-related experience. Participants received compensation at \$30/hour.

We employed a qualitative-dominant mixed-methods approach, prioritizing thematic analysis \cite{braunUsingThematicAnalysis2006a} of interview transcripts to identify key themes due to the small sample size ($N = 4$). Survey data were analyzed descriptively, focusing on individual shifts in perceptions and practices of CRP/CRT integration with and without LLM-based tools in AI literacy settings. Methodological triangulation ensured reliability. Given the exploratory nature of the pilot study, findings offer insights rather than generalizable conclusions.

\vspace{-4mm}
\section {Preliminary Findings}
\vspace{-3mm}

\subsection{AI Literacy in Schools: Teachers' Concerns and Practices}

Survey data revealed teachers' active use of AI tools, such as ChatGPT and Turnitin, for lesson planning, content creation, and student support. However, challenges remain in verifying AI-generated content accuracy, managing subscriptions, and addressing varied student comfort levels with technology. Participants show varied familiarity with AI tools, balancing concerns with exploring potential educational uses. P1, teaching in a virtual public school, expressed worries about students potentially misusing AI, asking, \textit{"How can we stop students from cheating all the time?"} This highlights broader concerns around academic integrity with AI access. Yet, there is recognition of the importance of teaching responsible AI usage, as P1 noted the necessity of guiding students to use AI constructively. For example, P1 used ChatGPT to simplify complex scientific texts and observed students creating concept maps with AI. P4 used LLM to generate multiple assignment examples from specified criteria for student selection. Despite these potential benefits, accuracy remains a concern: P1 noted occasional hallucinations in AI-generated explanations, and P4 highlighted the ongoing need to develop clearer AI usage guidelines and policies.

\vspace{-4mm}

\subsection{CRP in Schools: Resources, Support, and Challenges}

Interviews revealed varied levels of support for CRP across school environments. P2 indicated that although CRP is not explicitly mandated, it \textit{"is highly expected,"} with administrators observing and providing feedback. They also noted professional development opportunities, mentioning \textit{"they brought in a speaker and author who has written books on culturally relevant pedagogy."} In contrast, P4 remarked, \textit{"when I was in college I didn't hear anything about culturally responsive teaching. And that was 15 years ago,"} adding that current workplace discussions are \textit{"minimally, but not like at length."} P1 emphasized minimal support in previous experiences, noting, \textit{"in other schools that I have worked for, it was not required there really were not a whole lot of supports."} These highlighted significant variability in institutional support and available resources for CRP. Teachers also described challenges in effectively implementing CRP, notably the considerable time commitment required. P2 emphasized the difficulty of authentically knowing students beyond superficial traits, stating, \textit{"it really takes sitting down, knowing my students not just simply on skin color, but also socioeconomic background. It's a challenge because it takes time to do."} P1 shared similar concerns, initially questioning, \textit{"how would I have time to incorporate all of this information that is relevant to my students and their experiences?"} Authenticity emerged as another issue: P1 recounted experiences at schools where activities labeled as "culturally responsive" lacked genuine connection to students' lives. P4 acknowledged uncertainty in making meaningful connections, admitting, \textit{"I can just make assumptions. I guess I do my best."} Pre-survey results align with these perspectives. Participants reported moderate confidence (3 to 4 on a 5-point scale) in identifying and implementing CRP strategies, citing limited time and uncertainty about best practices as primary challenges.

\vspace{-4mm}

\subsection{CulturAIEd: Strengths and Opportunities}

Participants praised CulturAIEd for streamlining CRP, particularly through its integration of student demographics, content customization, rubric-based feedback. Therefore, using CulturAIEd was viewed as highly efficient. Post-surveys confirmed high satisfaction, with all participants reporting increased confidence in implementing CRP modifications. In specific, teachers' confidence in "Identifying opportunities for cultural responsiveness in learning activities" rose significantly from 3-5 in the pre-survey to all 5 in the post-survey.  P1 noted, "\textit{I hadn’t taken the time to consider culturally responsive practices, and it's making me realize it’s not so difficult to incorporate.}" P4 similarly expressed that the experience provided a helpful framework for designing culturally responsive activities. Identified opportunities for improvement include workflow enhancement, UI refinements, export functionality, and multilingual support.
\vspace{-3mm}
\begin{table}[h]
    \renewcommand{\arraystretch}{1.2}
    \fontsize{7pt}{7pt}\selectfont
    
    \begin{tabularx}{\textwidth}{>{\setlength\hsize{1\hsize}\raggedright}l|X|X}
        & \textbf{Aspect}                                                                                                           & \textbf{Supporting Quotes}                                                                                                                               \\ \hline
        \textbf{Strengths}     & CRP Integration: Facilitates culturally responsive pedagogy (CRP) with actionable insights.                      & \textit{“Extremely confident that CulturAIEd is very helpful… an enlightening tool”} (P2); \textit{“I do see myself using CulturAIEd… helpful for concrete examples”} (P1). \\ \cline{2-3}
                      & Streamlined Demographics \& Modifications: Simplifies integrating student demographics and activity adjustments. & \textit{“Demographics information was straightforward… nice to see integrated”} (P2); \textit{“Solved the problem of time and research needed for CRP”} (P1).     \\  \cline{2-3}
                      & Rubric Feature: Provides structured feedback for culturally responsive design.                                   & \textit{“The rubric broke down areas teachers might not consider”} (P2); \textit{“Liked the grading… feedback on all categories”} (P4).                           \\ \cline{2-3}
                      & Post-Survey Results: High confidence in CRP modifications; satisfaction with outputs.                            & All participants rated themselves 5/5 in post-surveys; 3/4 strongly agreed CulturAIEd saved time and inspired new ideas.                             \\  \hline
        \textbf{Opportunities} & Workflow Enhancements: Requests for automation and reduced manual input.                                         & \textit{“Automatically add information to a certain part”} (P1); \textit{“Integrate content with minimal copy/paste”} (P4).                                       \\ \cline{2-3}
                      & User Interface (UI) Improvements: Need for a more intuitive interface.                                           & \textit{“The UI could use a little work”} (P1).                                                                                                          \\ \cline{2-3}
                      & Exporting Capabilities: Demand for direct export to common formats (e.g., Google Docs).                          & \textit{“Export to Google Docs or Word would be helpful”} (P1); \textit{“Output in a document format… probably easiest”} (P4).                                    \\ \cline{2-3}
                      & Multi-Language Support                                            & \textit{“More explicit multi-language support”} (P3).                                                                                                    
    \end{tabularx}
    \caption{Strength and Opportunities of CulturAIEd}
    \vspace{-10mm}
    \label{tab:CulturAIEd-strength-opportunities}
    \end{table}

\section{Discussion \& Future Work}
\vspace{-3mm}

Results indicated a notable positive shift in teacher perceptions and practices regarding CRP after using CulturAIEd. We framed the study as a preliminary exploration to gather qualitative feedback. Accordingly, the improvements in confidence and efficiency should be interpreted as promising trends rather than conclusive evidence of efficacy. Future larger-scale studies (with control comparisons) will be needed to test CulturAIEd’s impact more rigorously.

Despite these positive changes, it is crucial to acknowledge the risks that come with using LLMs in this context. LLMs learn from vast datasets that may contain biases or stereotypes. If not carefully mitigated, the model could inadvertently produce content that reinforces simplistic cultural clichés or even offensive stereotypes, thereby undermining the culturally relevant pedagogy we seek to advance. Existing literature also warns against the risk of "routinizing" CRP \cite{leonardComplexitiesCulturallyRelevant2009}. Tools driven by LLMs, like CulturAIEd, could potentially become viewed as complete solutions rather than foundational resources. Although these tools support teaching through brainstorming, summarizing, and targeting, they cannot replace teachers' contextual understanding. Effective training is essential to help teachers adapt and refine AI suggestions for diverse student backgrounds.

Future research should examine the long-term impacts of LLM-supported tools like CulturAIEd on teacher practices and student outcomes in CRP across different subjects. Longitudinal studies could provide deeper insights into the sustained integration of CRP in lesson planning and classroom implementation. Moreover, the current sample size of four participants is too small, and further data collection from a broader range of contexts is essential to enhance the reliability and generalizability of the findings. Improvements to CulturAIEd, such as an enhanced user interface, better export options, and multi-language capabilities, should be prioritized to serve diverse educators effectively. Lastly, integrating such tools within professional development programs to encourage teacher reflection and refinement beyond template-based outputs represents another promising research direction.

\vspace{-4mm}

\bibliographystyle{splncs04}
\bibliography{reference}

\begin{thebibliography}{10}
\providecommand{\url}[1]{\texttt{#1}}
\providecommand{\urlprefix}{URL }
\providecommand{\doi}[1]{https://doi.org/#1}

\bibitem{AI4K12}
{{AI4K12}}. https://ai4k12.org/

\bibitem{ProlificEasilyCollect}
Prolific, \url{https://www.prolific.com/}

\bibitem{StreamlitFasterWay2021}
Streamlit, \url{https://streamlit.io/}

\bibitem{CulturallyResponsiveTeaching2019}
Culturally {{Responsive Teaching Checklist}}. https://reimaginingmigration.org/resource-items/cultural-responsive-teaching-checklist/ (Jan 2019)

\bibitem{aronsonTheoryPracticeCulturally2016a}
Aronson, B., Laughter, J.: The {{Theory}} and {{Practice}} of {{Culturally Relevant Education}}: {{A Synthesis}} of {{Research Across Content Areas}}  \textbf{86}(1),  163--206 (2016). \doi{10.3102/0034654315582066}

\bibitem{baytak2024content}
Baytak, A.: The content analysis of the lesson plans created by chatgpt and google gemini. Research in Social Sciences and Technology  \textbf{9}(1),  329--350 (2024)

\bibitem{braunUsingThematicAnalysis2006a}
Braun, V., Clarke, V.: Using thematic analysis in psychology. Qualitative Research in Psychology  \textbf{3}(2),  77--101 (Jan 2006). \doi{10.1191/1478088706qp063oa}

\bibitem{byrdDoesCulturallyRelevant2016}
Byrd, C.M.: Does {{Culturally Relevant Teaching Work}}? {{An Examination}} from {{Student Perspectives}}  \textbf{6}(3) (2016). \doi{10.1177/2158244016660744}

\bibitem{casal-oteroAILiteracyK122023a}
Casal-Otero, L., Catala, A., Fernández-Morante, C., Taboada, M., Cebreiro, B., Barro, S.: {{AI Literacy}} in {{K-12}}: {{A Systematic Literature Review}}  \textbf{10}. \doi{10.1186/s40594-023-00418-7}

\bibitem{deeCausalEffectsCultural2017}
Dee, T., Penner, E.: The {{Causal Effects}} of {{Cultural Relevance}}: {{Evidence}} from an {{Ethnic Studies Curriculum}}  \textbf{54(1)}, ~127 (2017)

\bibitem{drugaLandscapeTeachingResources2022}
Druga, S., Otero, N., Ko, A.J.: The {{Landscape}} of {{Teaching Resources}} for {{AI Education}}. In: Proceedings of the 27th {{ACM Conference}} on on {{Innovation}} and {{Technology}} in {{Computer Science Education Vol}}. 1. pp. 96--102. ACM (2022). \doi{10.1145/3502718.3524782}

\bibitem{eguchiContextualizingAIEducation2021}
Eguchi, A., Okada, H., Muto, Y.: Contextualizing {{AI Education}} for {{K-12 Students}} to {{Enhance Their Learning}} of {{AI Literacy Through Culturally Responsive Approaches}}  \textbf{35}(2),  153--161 (2021). \doi{10.1007/s13218-021-00737-3}

\bibitem{gayCulturallyResponsiveTeaching2000}
Gay, G.: Culturally {{Responsive Teaching}}: {{Theory}}, {{Research}}, and {{Practice}}. Teachers College Press (2000)

\bibitem{hammondCulturallyResponsiveTeaching2015}
Hammond, Z., Jackson, Y.: Culturally {{Responsive Teaching}} and the {{Brain}}  (2015)

\bibitem{johnsonCulturallyRelevantPedagogy2020}
Johnson, A., Elliott, S.: Culturally {{Relevant Pedagogy}}: {{A Model To Guide Cultural Transformation}} in {{STEM Departments}}  \textbf{21}(1),  21.1.35 (2020). \doi{10.1128/jmbe.v21i1.2097}

\bibitem{10.1007/978-3-031-64315-6_49}
Laak, K.J., Aru, J.: Generative ai in k-12: Opportunities for learning and utility for teachers. In: Olney, A.M., Chounta, I.A., Liu, Z., Santos, O.C., Bittencourt, I.I. (eds.) Artificial Intelligence in Education. Posters and Late Breaking Results, Workshops and Tutorials, Industry and Innovation Tracks, Practitioners, Doctoral Consortium and Blue Sky. pp. 502--509. Springer Nature Switzerland, Cham (2024)

\bibitem{ladson-billingsTheoryCulturallyRelevant1995}
Ladson-Billings, G.: Toward a {{Theory}} of {{Culturally Relevant Pedagogy}}  \textbf{32}(3),  465--491 (1995). \doi{10.3102/00028312032003465}

\bibitem{leonardComplexitiesCulturallyRelevant2009}
Leonard, J., Napp, C., Adeleke, S.: The {{Complexities}} of {{Culturally Relevant Pedagogy}}: {{A Case Study}} of {{Two Secondary Mathematics Teachers}} and {{Their ESOL Students}}  \textbf{93}(1),  3--22 (2009). \doi{10.1353/hsj.0.0038}

\bibitem{li2025unseen}
Li, H., Xiao, R., Nieu, H., Tseng, Y.J., Liao, G.: “from unseen needs to classroom solutions”: Exploring ai literacy challenges \& opportunities with project-based learning toolkit in k-12 education. In: Proceedings of the AAAI Conference on Artificial Intelligence. vol.~39, pp. 29145--29152 (2025)

\bibitem{longWhatAILiteracy2020a}
Long, D., Magerko, B.: What is {{AI Literacy}}? {{Competencies}} and {{Design Considerations}}. In: Proceedings of the 2020 {{CHI Conference}} on {{Human Factors}} in {{Computing Systems}}. pp. 1--16. ACM (2020). \doi{10.1145/3313831.3376727}

\bibitem{mette2016teachers}
Mette, I.M., Nieuwenhuizen, L., Hvidston, D.J.: Teachers' perceptions of culturally responsive pedagogy and the impact on leadership preparation: Lessons for future reform efforts. International Journal of Educational Leadership Preparation  \textbf{11}(1), ~n1 (2016)

\bibitem{moorhousePreserviceLanguageTeachers2025}
Moorhouse, B.L., Ho, T.Y., Wu, C., Wan, Y.: Pre-service {{Language Teachers}}’ {{Task-specific Large Language Model Prompting Practices}} p. 00336882251313701. \doi{10.1177/00336882251313701}

\bibitem{royChatGPTLessonPreparation}
Roy, P., Staunton, R., Poet, H.: {{ChatGPT}} in lesson preparation - {{A Teacher Choices Trial}}. \doi{10.1186/ISRCTN13420346}

\bibitem{samuels2018exploring}
Samuels, A.J.: Exploring culturally responsive pedagogy: Teachers' perspectives on fostering equitable and inclusive classrooms. Srate Journal  \textbf{27}(1),  22--30 (2018)

\bibitem{tseng2024activeai}
Tseng, Y.J., Yadav, G., Hou, X., Wu, M., Chou, Y.S., Chen, C.C., Wu, C.C., Chen, S.G., Lin, Y.J., Liao, G., et~al.: Activeai: The effectiveness of an interactive tutoring system in developing k-12 ai literacy. In: European Conference on Technology Enhanced Learning. pp. 452--467. Springer (2024)

\end{thebibliography}

\end{document}